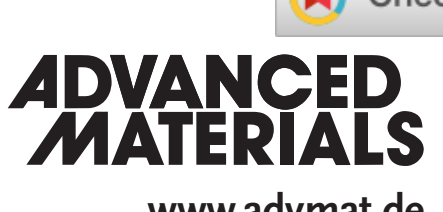

# Magnetic-Field Tunable Intertwined Checkerboard Charge Order and Nematicity in the Surface Layer of $Sr_2RuO_4$

*Carolina A. Marques, Luke C. Rhodes, Rosalba Fittipaldi, Veronica Granata, Chi Ming Yim, Renato Buzio, Andrea Gerbi, Antonio Vecchione, Andreas W. Rost, and Peter Wahl**

**In strongly correlated electron materials, the electronic, spin, and charge degrees of freedom are closely intertwined. This often leads to the stabilization of emergent orders that are highly sensitive to external physical stimuli promising opportunities for technological applications. In perovskite ruthenates, this sensitivity manifests in dramatic changes of the physical properties with subtle structural details of the $RuO_6$ octahedra, stabilizing enigmatic correlated ground states, from a hotly debated superconducting state via electronic nematicity and metamagnetic quantum criticality to ferromagnetism. Here, it is demonstrated that the rotation of the $RuO_6$ octahedra in the surface layer of $Sr_2RuO_4$ generates new emergent orders not observed in the bulk material. Through atomic-scale spectroscopic characterization of the low-energy electronic states, four van Hove singularities are identified in the vicinity of the Fermi energy. The singularities can be directly linked to intertwined nematic and checkerboard charge order. Tuning of one of these van Hove singularities by magnetic field is demonstrated, suggesting that the surface layer undergoes a Lifshitz transition at a magnetic field of ≈32T. The results establish the surface layer of $Sr_2RuO_4$ as an exciting 2D correlated electron system and highlight the opportunities for engineering the low-energy electronic states in these systems.**

## 1. Introduction

The physical properties of strongly correlated electron materials often vary dramatically with comparatively modest external stimuli,[1] evidenced, for example, by magnetic-field driven metamagnetic transitions,[2,3] doping-induced metal-to-insulator transitions[4] and superconductivity[5] as well as a surprising sensitivity to uniaxial strain.[6] The changes in physical properties are usually accompanied by tiny structural distortions often reflecting, or even inducing, the lower symmetry of the new electronic states. This sensitivity of physical properties is exemplified in the perovskite ruthenates. The members of the Ruddlesden–Popper series of strontium ruthenate, $Sr_{n+1}Ru_nO_{3n+1}$, exhibit an exceptional variety of ground states ranging from superconductivity for $n$ = 1[7] via materials with a rich metamagnetic phase diagram for $n$ = 2[8] and 3[9] to bulk ferromagnetism for $n \to \infty$.[10] This wide variety of properties is intimately linked to small structural distortions of the $RuO_6$ cage. Already in the superconductor $Sr_2RuO_4$, with $n$ = 1, a soft phonon mode associated with the octahedral rotation is observed.[11] Isoelectronic substitution of Sr by Ca results in a rich phase diagram with metallic and ferromagnetic phases until an antiferromagnetically ordered Mott insulator is reached, where the dominant change to the material is a small rotation and tilting of the $RuO_6$ cage.[1,4] Understanding the impact of that rotation on the electronic structure therefore provides a key to controlling electronic and physical properties of perovskite ruthenates. A prerequisite to link physical properties to details of the octahedral rotation is an understanding of the low-energy electronic structure. We here establish the surface layer of $Sr_2RuO_4$ as a clean and well-controlled 2D model system where this can be achieved and explored.

The bulk material of $Sr_2RuO_4$ has a tetragonal unit cell with undistorted $RuO_6$ octahedra[12,13] aligned with the high-symmetry directions of the crystal (left panel **Figure 1**a). It is a superconductor with a transition temperature $T_c \approx 1.5$K,[7] and is known to exhibit strong electron correlations, evidenced by effective masses of the bands between 6 and $17m_e$ in the normal state.[14] Its bulk Fermi surface has been established by quantum oscillations[15] and confirmed by angle-resolved photoemission spectroscopy (ARPES).[14,16,17]

C. A. Marques, Dr. L. C. Rhodes, Prof. C. M. Yim, Dr. A. W. Rost, Prof. P. Wahl
School of Physics and Astronomy
University of St Andrews
North Haugh, St Andrews, Fife KY16 9SS, UK
E-mail: wahl@st-andrews.ac.uk

Dr. R. Fittipaldi, Dr. A. Vecchione
CNR-SPIN
UOS Salerno
Via Giovanni Paolo II 132, Fisciano I-84084, Italy

Dr. V. Granata
Dipartimento di Fisica "E. R. Caianiello" Universitá di Salerno
Fisciano, Salerno I-84084, Italy

Dr. R. Buzio, Dr. A. Gerbi
CNR-SPIN
Corso F.M. Perrone 24, Genova 16152, Italy

Dr. A. W. Rost
Max-Planck-Institute for Solid State Research
Heisenbergstr. 1, 70569 Stuttgart, Germany

The ORCID identification number(s) for the author(s) of this article can be found under https://doi.org/10.1002/adma.202100593.

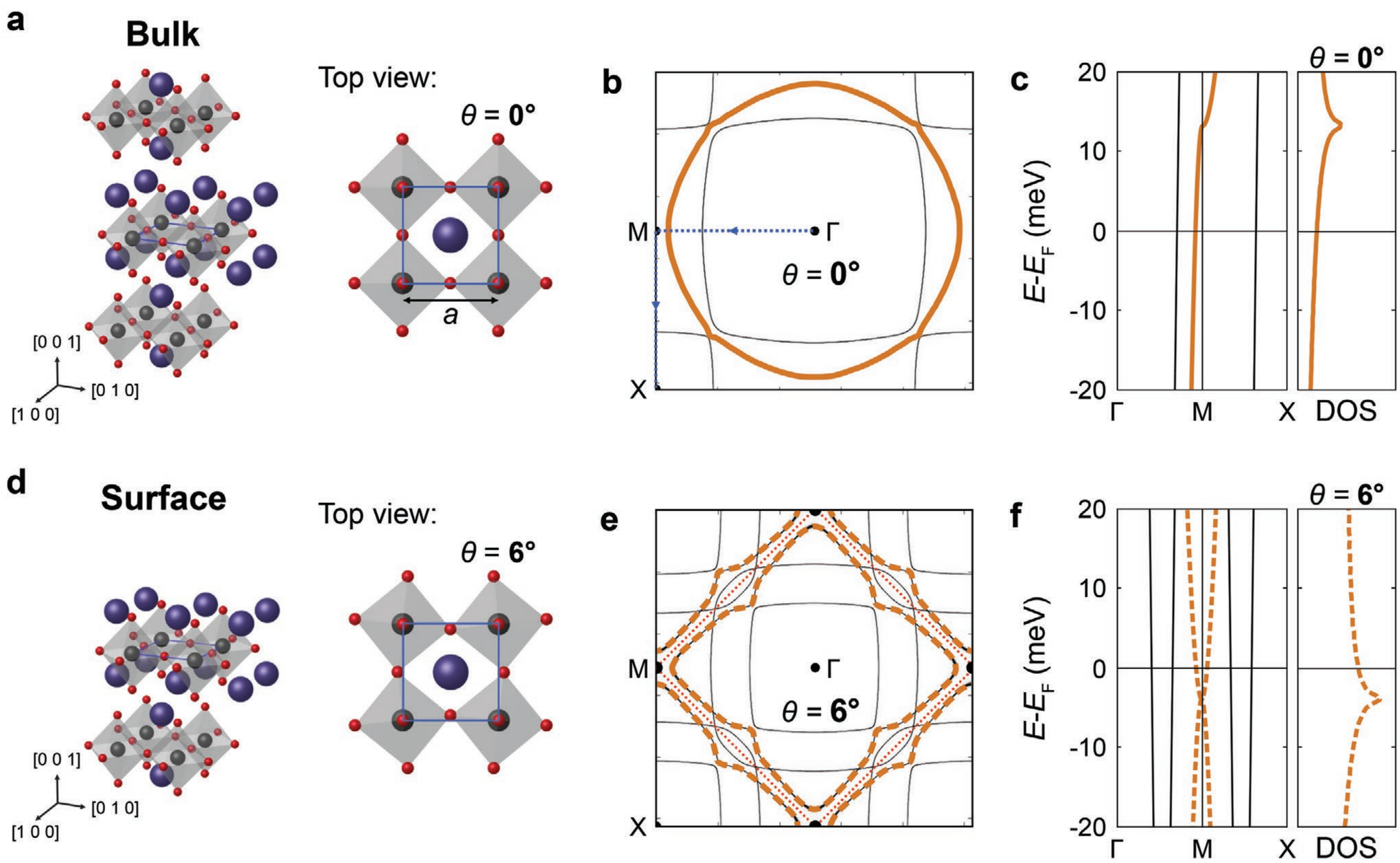

**Figure 1.** Electronic structure of $Sr_2RuO_4$ with and without octahedral rotation. a) Structural model of the bulk of $Sr_2RuO_4$ (black: Ru, red: O, purple: Sr atoms). Right: top view, with the unit cell of lateral size *a* indicated by a blue square. The octahedra have the same orientation at each lattice site, with an angle of $\theta = 0°$. b) Fermi surface of $Sr_2RuO_4$ with $\theta = 0°$[18] (yellow line: $d_{xy}$ band, black lines: $d_{xz}/d_{yz}$ bands). c) Electronic structure of bulk $Sr_2RuO_4$. The $d_{xy}$ band exhibits a van Hove singularity (vHs) at the M point above $E_F$, which results in a peak in the density of states (DOS, right). d) Structural model of the surface of $Sr_2RuO_4$ with $\theta = 6°$. The rotation is more apparent in the top view shown on the right and leads to a doubling of the unit cell. The unit cell with $\theta = 0°$ is shown as a blue square for comparison. e) Fermi surface with $\theta = 6°$ (yellow dashed line: $d_{xy}$ band; see Section S4, Supporting Information for details). The surface Brillouin zone is indicated by dotted red lines. f) Electronic structure corresponding to (e). The vHs at the M-point is pushed below $E_F$ with the octahedral rotation. The DOS is shown on the right, where the peak occurs below $E_F$.

In Figure 1b, we show its Fermi surface: The $d_{xy}$ band (yellow line) has a van Hove singularity (vHs) at the M-point of the Brillouin zone (BZ) at about 14 meV above $E_F$ (Figure 1c).[19] Upon cleaving, the exposed surface layer exhibits a 6° rotation of the $RuO_6$ octahedra[20] (Figure 1d). In bulk $Sr_2RuO_4$, a similar octahedral rotation occurs on substitution of Sr by Ca quickly suppressing superconductivity.[4] Early STM studies[21,22] suggest that the rotation also suppresses superconductivity in the surface layer. The rotation further leads to orbital-dependent renormalizations close to the Fermi energy.[23,24] This rotation doubles the size of the unit cell and leads to a reconstruction of the Fermi surface (Figure 1d), and a shift of the vHs below $E_F$ (Figure 1e).[17] The vHs leads to a peak in the density of states (DOS). In Figure 1c,e) we show schematically the DOS of the unreconstructed (solid line) and reconstructed (dashed line) electronic structure in comparison. The energy of the vHs is seen to be highly sensitive to structural details of the $RuO_6$ octahedra. Due to the small interlayer coupling in $Sr_2RuO_4$, the surface reconstruction provides an opportunity to establish the influence of small structural distortions on the electronic structure in a very clean and chemically homogeneous 2D system, and hints at what the leading instabilities of the bulk material are. We here demonstrate that the reconstructed surface of $Sr_2RuO_4$ by itself constitutes a strongly correlated system with its own unique electronic properties: checkerboard charge order, nematicity and four van Hove singularities (vHss) within a few millielectronvolts of the Fermi energy. We can link these phenomena through a phenomenological tight-binding model based on the bulk electronic structure and incorporating these emergent orders which yields a density of states in excellent qualitative agreement with tunnelling spectra. We demonstrate magnetic-field tuning of one of the vHss, and extrapolate from our measurements that the surface layer undergoes a Lifshitz transition at 32T.

## 2. Results

### 2.1. Checkerboard Charge Order

**Figure 2**a shows a topographic image of $Sr_2RuO_4$ at a bias voltage $V = 5$ mV, recorded at a temperature of 76 mK, showing a SrO-terminated surface[25,26] with atomic resolution and demonstrating a low concentration of point defects (less than 0.1%). We find defects at the Ru site with two distinct orientations due to the octahedral rotation in the surface layer. The Fourier transformation of the topography (upper inset in Figure 2a),

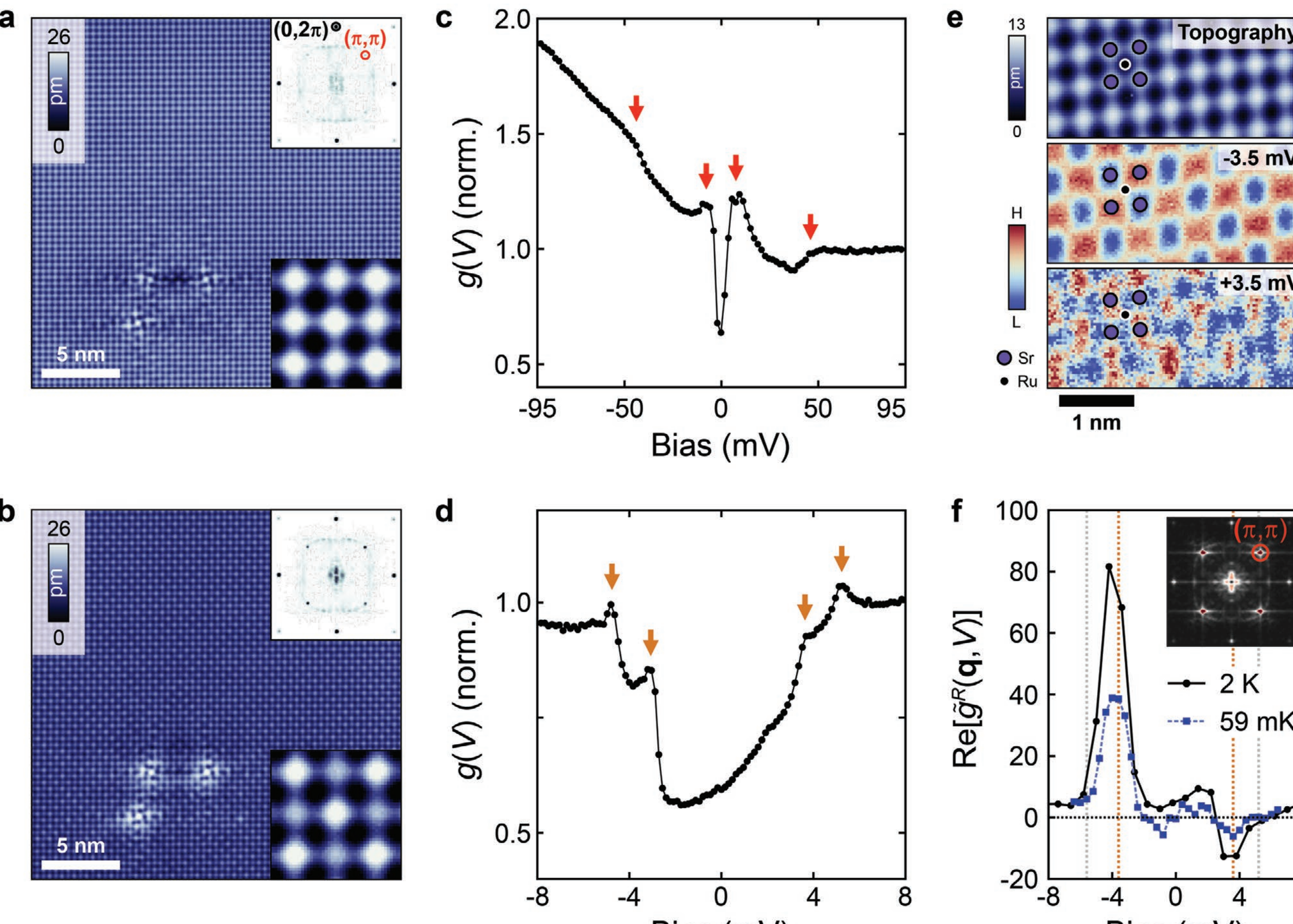

**Figure 2.** Checkerboard charge order. a) Topography taken at $V = +5$ mV, showing the Sr square lattice ($I_{set} = 50$ pA). Lower inset: enlarged topography. Upper inset: Fourier transformation with Bragg peaks at (0, $2\pi$) (black circle) and ($2\pi$, 0). Peaks at ($\pi$, $\pi$) (red circle) and ($-\pi$, $\pi$) coincide with the periodicity of the surface reconstruction (reciprocal lattice vectors in units of $1/a$). b) Topography at $V = -5$ mV, showing a clear checkerboard ($I_{set} =$ 50pA), shown in more detail in the lower inset. Upper inset: Fourier transformation showing increase of intensity at ($\pi$, $\pi$). c) Tunneling spectrum $g(V)$ measured at $T = 76$ mK ($V_{set} = 100$ mV, $I_{set} = 265.2$ pA, $V_L = 1.75$ mV). d) High resolution $g(V)$ spectrum around $E_F$ with a gap and four peaks indicated by yellow arrows ($V_{set} = 8$ mV, $I_{set} = 500.2$ pA, $V_L = 155$ μV, $T = 56$ mK). e) Top: Topography with a model indicating the positions of the Sr atoms. Bottom: Real-space $g(\mathbf{r},V)$ maps at $V = -3.5$ mV and $V = +3.5$ mV recorded simultaneously with the topography. At $-3.5$ mV, a strong checkerboard charge order is observed which has opposite phase at +3.5 mV ($T = 59$ mK, $V_{set} = 7.0$ mV, $I_{set} = 250$ pA, $V_L = 495$ μV). f) Energy dependence of the phase-referenced Fourier transformation $\tilde{g}^R(\mathrm{q},V)$ at $\mathrm{q}_{ckb} = (\pi,\pi)$ obtained from maps taken at $T = 2$ K and 59 mK. Two peaks at $-3.5$ mV and +3.5 mV with opposite phase and a width of ≈1 mV can be seen (see Section S3, Supporting Information for details). Vertical dotted lines indicate the position of the four peaks observed in the average $g(\mathrm{r},V)$ spectrum at 59 mK. Insert: Fourier transformation $\tilde{g}(\mathrm{q},V)$ for the $T = 2$ K map, $\mathrm{q}_{ckb}$ is indicated by a red circle.

shows the presence of quasi-particle interference (QPI), consistent with previous reports,[27] as well as Bragg peaks at (0, $2\pi$) and ($2\pi$, 0) due to the Sr lattice, and peaks with a low intensity at ($\pi$, $\pi$) corresponding to the periodicity of the surface reconstruction. When recording topographic STM images at bias voltages of −5 mV, Figure 2b, the appearance changes significantly. The image contrast is now dominated by the presence of a strong modulation of the charge density (lower inset in Figure 2b) which corresponds to a large increase in the intensity of the ($\pi$, $\pi$) peaks in the Fourier transformation. This modulation has been observed previously at the surface of $Sr_2RuO_4$,[25] as well as at that of $Sr_3Ru_2O_7$.[28] However, the octahedral rotation due to the surface reconstruction cannot explain the charge modulation: adjacent Sr and Ru sites are in an equivalent structural environment that can be transformed into each other through a symmetry operation—a mirror operation for Ru and a 90° rotation for Sr. Detailed studies of the surface structure by low-energy electron diffraction (LEED) do not reveal any additional reconstruction of the surface layer apart from the octahedral rotation,[29] leaving the origin of the checkerboard charge order as an open question. We note that the checkerboard charge order is most prominent for small bias voltages, suggesting an electronic origin (see Section S2, Supporting Information). A typical differential conductance spectrum $g(V)$ in the range ±95 mV is presented in Figure 2c. Here, kink- and gap-like features are observed at ±40 meV and ±5 meV, respectively. The latter is associated with a reduction of the differential conductance by almost 30% in relation to the value at 95 mV. The general shape of the spectrum, as well as the absence of a superconducting gap, is in agreement with previous reports.[21,22,27] However, from high-resolution spectra acquired at temperatures below 100 mK, we find that this gap-like feature actually exhibits four well defined peaks in the differential conductance within ±5 meV of the Fermi energy (Figure 2d).

The additional modulation of the charge density leads to pronounced signatures in spectroscopic maps: while topographic images obtained at positive bias voltage show predominantly the Sr lattice (top panel in Figure 2e), differential conductance maps exhibit a clear checkerboard charge modulation superimposed to this lattice at bias voltages close to $E_F$ (second and third panel in Figure 2e). The checkerboard exhibits a contrast inversion across the Fermi energy. We plot in Figure 2f the energy dependence of the intensity of the checkerboard analyzed through a phase-referenced Fourier transformation (PR-FT, see Section S3C, Supporting Information), which provides information about the amplitude and phase of the modulation. The amplitude of the checkerboard exhibits a pronounced maximum at −3.5 mV with a width of only about 1mV and a weaker maximum at +3.5 mV of similar width but with opposite phase. Both maxima occur at the same energy as the energy of the inner-most peaks in the tunneling spectrum (yellow dotted lines in Figure 2f). The phase change is consistent with what one would expect for a charge density wave.[30]

### 2.2. Nematicity

The Sr-centred checkerboard charge order and nematicity are intimately linked in the surface layer through the octahedral rotation. The octahedral rotation itself preserves $C_4$ symmetry and does not give rise to an additional charge modulation or nematicity. However, the occurrence of a Sr-centred checkerboard, as we observe experimentally, and nematicity are equivalent: due to the checkerboard charge order centred at the Sr sites and the octahedral rotation, the oxygen atoms in the Ru plane become inequivalent between the horizontal, [10], and vertical, [01], directions. As shown in **Figure 3**a, the rotation means that the oxygen atoms connecting Ru atoms in the [01] (vertical) direction (colored in red) are closer to Sr atoms with a decreased charge density (shown dark), whereas oxygen atoms in the [10] (horizontal) direction (shown orange) are closer to Sr atoms with increased charge density. This leads to different hopping amplitudes across those oxygen atoms, indicated in Figure 3a by the yellow and red dashed lines, resulting in nematicity of the electronic states and a reduction from $C_4$ symmetry to $C_2$ symmetry (Figure 3a). The converse holds true as well, nematicity on the oxygen sites (indicated by different coloured oxygen atoms in Figure 3a) results in their inequivalence and hence checkerboard charge order on the Sr sites – highlighting the equivalence of the two orders. Therefore, while the octahedral rotation itself gives rise to neither nematicity nor checkerboard charge order, the two become intimately related through the octahedral rotation—so if one occurs, the other will too. This emergent nematicity is confirmed experimentally through unidirectional modulations with atomic periodicity (Figure 3b,c) and anisotropy of the low-$\mathbf{q}$ quasi-particle interference (Figure 3d–i). The atomic scale symmetry breaking reveals that for changes in the bias voltage $V$ by about 1 mV the atomic-scale unidirectional periodicity changes direction between the high-symmetry directions [10] and [01] implying a small characteristic energy scale of the nematicity. This is also confirmed in the intensity of the atomic peaks (Figure S5, Supporting Information). The symmetry breaking of long-wavelength quasi-particle interference is shown in a real-space map around four defects in Figure 3d–i (defects marked by black crosses) and reveals a change from predominant quasiparticle scattering along [10] to scattering along [01] and back as a function of energy.

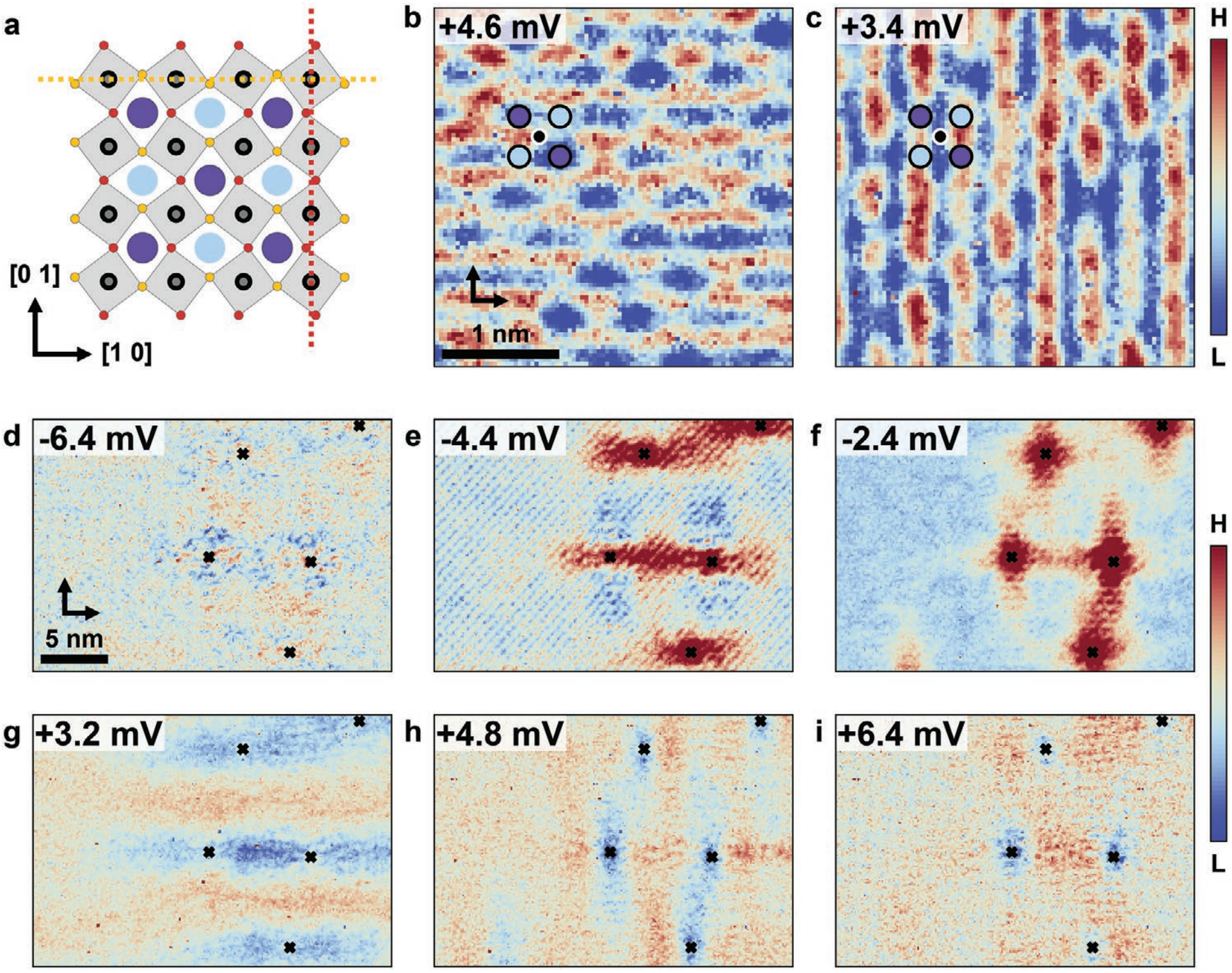

**Figure 3.** Nematicity and the equivalence of checkerboard charge order with $C_4$ symmetry breaking. a) Model of the surface atomic structure with the checkerboard charge order on the Sr atoms (purple and light blue circles). The charge order on the Sr lattice combined with the octahedral rotations leads to a broken $C_4$ symmetry, due to which oxygen atoms along the [10] and [01] directions are in an inequivalent environment. This results in inequivalent hopping amplitudes along the horizontal (yellow) and vertical (red) dashed lines. The color of the oxygen atoms represents whether they are closer to a light blue Sr atom (yellow) or a purple Sr atom (red). b,c) Nematicity in the atomic scale charge modulations ($T$ = 1.8 K, $V_{set}$ = 7.8 mV, $I_{set}$ = 500 pA, $V_L$ = 370 µV). d–i) Real-space images showing directional quasi-particle interference near defects on a longer length scale, the defects are marked by black crosses ($T$ = 56 mK, $V_{set}$ = 6.4 mV, $I_{set}$ = 225 pA, $V_L$ = 398 µV).

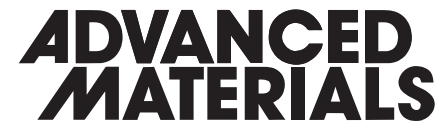

We note that neither nematicity nor checkerboard charge order are captured by the surface structure determined by $I(V)$-LEED.[29] Nor does $I(V)$-LEED or our STM data provide any evidence for an orthorhombicity of the surface layer which could give rise to a lowered symmetry.

### 2.3. Tight-Binding Model for Surface Electronic Structure

To understand the microscopic consequences nematicity and the observed charge modulation have on the electronic structure of the surface layer, we have developed a minimal tight-binding model for the band structure at the surface of $Sr_2RuO_4$. This model was generated by taking a tight-binding model of the bulk of $Sr_2RuO_4$,[18] and then altering the position of the $d_{xy}$ vHs such that it is shifted toward the Fermi level (see Section S4, Supporting Information for details). We then include the emergent orders at a phenomenological level. The surface reconstruction and the checkerboard charge order (Figure 2) are accounted for by including a weak intraband hybridization ($\Delta_{\mathrm{hyb}}$). The nematicity and inequivalence of the [10] and [01] lattice directions (Figure 3) are included through a nematic order parameter $\Delta_{\mathrm{nem}} = \delta_{\mathrm{nem}}(\cos(k_x) - \cos(k_y))$, which we apply only to the $d_{xy}$ orbital. A similar order parameter has previously been introduced to describe nematicity in $Sr_3Ru_2O_7$.[31] The full Hamiltonian is then

$$H(\mathbf{k}) = \begin{pmatrix} H_{\mathrm{Ru}}(\mathbf{k}) + \Delta_{\mathrm{nem}}\hat{I}_{xy} & \Delta_{\mathrm{hyb}}\hat{I} \\ \Delta_{\mathrm{hyb}}\hat{I} & H_{\mathrm{Ru}}(\mathbf{k}+\mathbf{Q}) + \Delta_{\mathrm{nem}}\hat{I}_{xy} \end{pmatrix} \quad (1)$$

where $H_{\mathrm{Ru}}(\mathbf{k})$ is the Hamiltonian for the ruthenium 4d $t_{2g}$ bands in the unreconstructed Brillouin zone,[18] and $\mathbf{Q} = (\pi, \pi)$ accounts for the doubling of the unit cell (for details see Section S4, Supporting Information). The inclusion of the emergent orders leads to the formation of four vHss around the M point, shown in **Figure 4**a which shows the band structure obtained from Equation (1). We follow the notation by van Hove[32] to label the four vHss: nematicity results in two saddle points, $\mathcal{S}1$ and $\mathcal{S}2$, on the high symmetry axis at the M point, whereas hybridization leads to a partial gap with a band maximum $\mathcal{M}$ and an additional saddle point around $E_F$. The resulting complex low-energy electronic structure around the M point is shown in a 3D representation in Figure 4b. The partial drop in the $g(V)$ spectrum around $E_F$, observed in Figure 2d, is naturally explained by the hybridization of the Ru $d_{xy}$-bands caused by the doubling of the unit cell. The four singularities lead to maxima in the density of states (see Figure 4c) in excellent agreement with the structure of the low-energy $g(V)$ spectrum. Differences, such as the magnitude of the drop, are likely a consequence of tunneling matrix element effects, which are neglected in the calculation.

### 2.4. Magnetic-Field Tuning of the Electronic Structure

The presence of multiple vHss in the immediate vicinity of the Fermi energy offers the opportunity to drive the material through a Lifshitz transition using magnetic field and Zeeman splitting of the bands. We investigate the influence of an applied magnetic field, focusing on the behaviour of the most prominent vHs, $\mathcal{M}$, at −3.4 mV. Figure 4d shows tunneling spectra recorded in magnetic fields from 0 T to 13.4 T, applied perpendicular to the surface, at $T = 76$ mK. The spectra reveal a clear splitting of the vHs with increasing magnetic field. In Figure 4e, the field-dependence of the energy of the vHs reveals a linear behavior as expected for a Zeeman-like splitting. We find the slope to be +0.10 mV $T^{-1}$ for the peak moving toward the Fermi energy and −0.07 mV $T^{-1}$ for the one moving away. The difference in slope can be attributed to an overall chemical potential shift of the $d_{xy}$ band of 17 μV $T^{-1}$. We find a $g$-factor of $g^* \approx 3$ (with the splitting $\Delta E = g^* \mu_B B_z$). This value is consistent with the known Wilson ratio $W = g^*/g = 1.5$ of bulk $Sr_2RuO_4$,[33,34] a surprisingly good agreement given that the Wilson ratio contains contributions from the whole Fermi surface, whereas we determine $g^*$ only for one band. From our experimental data, we can extrapolate that the vHs will cross the Fermi energy at a magnetic field of about 32 T. At this field, the surface layer is expected to undergo a Lifshitz transition. As the energy of the vHs is associated with the checkerboard charge order, once it is split in a magnetic field, the charge order becomes spin-polarized. We show in Figure 4f that indeed also the energy at which the checkerboard charge order appears most prominently shifts with the change in the energy of the vHs, demonstrating that the two are intimately linked. Our measurements demonstrate that the electronic structure of the reconstructed surface layer of $Sr_2RuO_4$ has all the ingredients for a field-tuned Lifshitz transition: 1) a vHs close to the Fermi energy, 2) the energy of the vHs can be tuned by magnetic field toward the Fermi energy, and 3) a Lifshitz transition of the electronic structure within reach of available magnetic fields.

## 3. Discussion

Our measurements show how the low-energy electronic structure of a 2D layer of $Sr_2RuO_4$ is impacted by octahedral rotations and stabilizes new emergent phases close to a magnetic-field induced Lifshitz transition. We provide a comprehensive description of the surface electronic structure through a phenomenological tight-binding model that incorporates the two orders we detect, nematicity and checkerboard charge order and reproduces all features we observe in our data. It captures the pronounced gap-like feature seen in tunneling spectra. The quick changes in the appearance of the nematicity in spectroscopic maps within a narrow energy interval are naturally explained by the vHss in the $k_x$- and $k_y$-directions becoming inequivalent. While checkerboard charge order and nematicity are intimately intertwined, it is not clear which of the two dominates, if either does. Possible mechanisms include 1) nematicity driven by electronic correlations, 2) charge order driven by nesting or a lattice distortion, 3) a cooperative effect of both, and 4) an antiferromagnetic order or fluctuations which stabilize the checkerboard order and nematicity. For the first three scenarios, one would expect a significant structural distortion accompanying these, as is seen for charge density waves or nematicity in other systems. A structural distortion in the surface layer of $Sr_2RuO_4$ beyond the octahedral rotation has not been reported.[29] In a scenario where magnetic order or fluctuations drive nematicity and charge order, one would still expect a structural distortion, though much smaller. Recently, evidence for nematicity in thin films of $Sr_2RuO_4$ has been reported, suggesting that it may not be limited to the surface layer.[35]

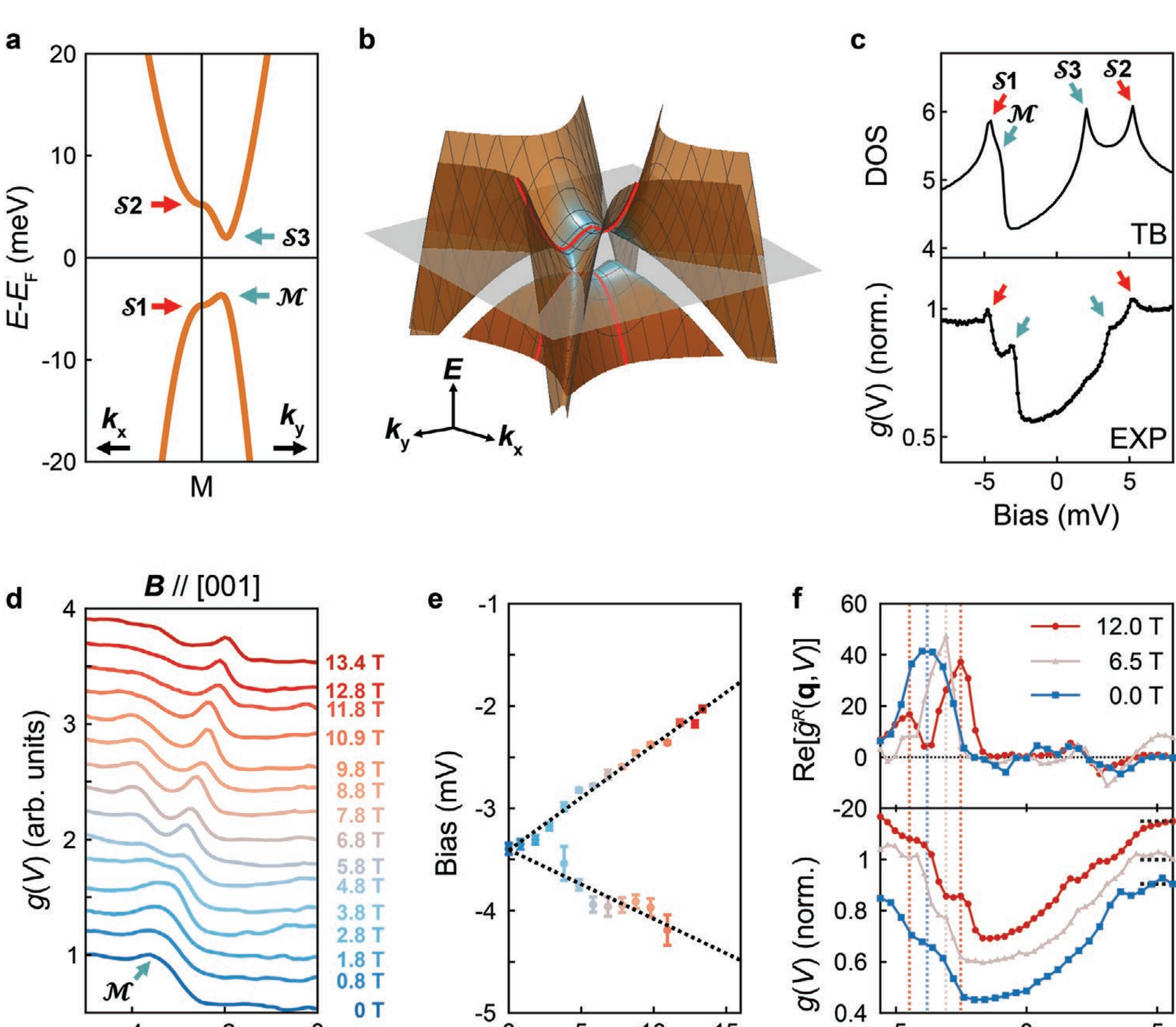

**Figure 4.** Tight-binding model and magnetic-field tuning of van Hove singularities. a,b) Band structure around M from the tight-binding model including an intraband hybridization potential and nematic order parameter. b) 3D band dispersion around M within 20 meV of $E_F$. Blue shaded regions indicate the locations of the vHs (labelled $\mathcal{S}1$, $\mathcal{S}2$, $\mathcal{S}3$, and $\mathcal{M}$ in (a)). Red lines: path shown in (a), the plane corresponds to $E_F$. c) Upper panel: Density of states from the model, with four vHss. Bottom: tunneling spectrum of Figure 2d with gap-like structure and four distinct peaks associated with the vHss. d) Tunneling spectra ($T = 76$ mK) in magnetic fields ($B||c$) from 0T to 13.4T ($V_{set} = 5$ mV, $I_{set} = 225$ pA, $V_L = 100$ µV). e) Peak position of $\mathcal{M}$ as a function of field extracted from fits (see Section S5, Supporting Information) revealing splitting and linear magnetic field dependence with $g^* \approx 3$ (error bars: 95% confidence intervals). f) Top: energy dependence of the Fourier peak at in the PR-FT at $B$ = 0T, 6.5T and 12T (see Section S3C, Supporting Information for details). Bottom: spatially averaged differential conductance spectra $g(V)$. The vertical dashed lines area guide to the eye, showing correspondence of features in $g(V)$ and the PR-FT. (Spectra are normalized and vertically shifted for clarity, the short dashed gray lines indicate $g(V)$ = 1). ($T = 76$ mK, $V_{set} = 5.6$ mV, $I_{set} = 225.2$ pA, $V_L = 300$ µV map at 6.5T; $T = 500$ mK, $V_{set} = 5.6$ mV, $I_{set} = 200.5$ pA, $V_L = 280$ µV for map at 12 T).

A comparison with $Sr_3Ru_2O_7$ reveals intriguing parallels: $Sr_3Ru_2O_7$ exhibits a similar (albeit larger) octahedral rotation, and the energy of the vHs as detected by ARPES is found at ≈4 meV below $E_F$,[36] close to the energy at which we find the dominant vHs in the surface layer of $Sr_2RuO_4$. Due to stronger correlations and a $g$-factor of $g^* \approx 14.6$ in $Sr_3Ru_2O_7$,[37] the vHs can be tuned to the Fermi energy at 8T, whereas in the surface layer of $Sr_2RuO_4$, we find a $g$-factor that is almost 4 times smaller. Consequently, the vHs is expected to reach the Fermi energy only at 32T. It remains to be seen whether the parallels go any further when the surface layer in $Sr_2RuO_4$ undergoes the Lifshitz transition. There are also important differences, though, including that the system we report here is strictly 2D which puts the criticality of the Lifshitz transition into a different universality class than what is expected for the bulk of $Sr_3Ru_2O_7$. Tunneling spectra recorded on $Sr_3Ru_2O_7$ do not reveal a clear shift of a peak as a function of magnetic field,[28] possibly because quantum fluctuations play a much larger role and already have influence on the line shape of the vHs in the density of states. Previous studies of the reconstructed surface have not been able to detect signatures of superconductivity,[21,22] except a recent study[38] where a gap-like structure in the vicinity of the Fermi energy has been attributed to superconductivity. Tunneling spectra showing clear signatures of superconductivity with a temperature and/or magnetic field dependence consistent with bulk $Sr_2RuO_4$[39–41] were either obtained from surfaces that were exposed to air leading effectively to a dirty surface much like in ARPES experiments probing the bulk electronic structure or on a different surface reconstruction than the one studied here. We do not observe any evidence for superconductivity even in tunneling spectra acquired at temperatures well below 100 mK. We have ensured that the spectroscopic resolution of our instrument is sufficient to detect superconducting gaps of materials with a similar transition temperature and gap sizes on the order of 200 μeV.[42,43] This indicates that the emergent electronic order suppresses superconductivity in the surface layer. This suppression may imply that the density of states of the $d_{xy}$ band around $E_F$, which becomes gapped out at the surface, and the fluctuations associated with the reconstruction play a crucial factor in superconducting pairing. This scenario naturally results in a competition of nematicity and the charge density modulations with superconductivity, reminiscent to what is found in other strongly correlated electron systems.[5,44] Understanding the leading instability in this highly debated material therefore undoubtedly will have to account for this susceptibility toward density wave formation as observed in other unconventional superconductors.

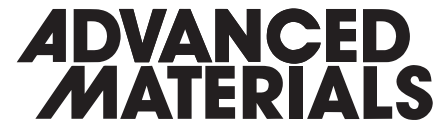

## 4. Conclusions

We show that the surface layer of $Sr_2RuO_4$ provides a 2D model system to study the intricate structure–property relationships of a strongly correlated electron system. We demonstrate the equivalence of checkerboard charge order and nematicity in this system, and find that the reconstructed electronic structure leads to four vHss within 5 meV of the Fermi energy. Magnetic-field tuning of one of these vHs implies that the surface layer can be used as a well-controlled test system to study magnetic-field induced Lifshitz transitions, enabling detailed comparison with microscopic theories. Because the emergent surface phase is strictly 2D, limited to the surface layer, all relevant information about the electronic states is accessible spectroscopically. Magnetic-field tuned Lifshitz transitions have been proposed to be at the heart of the quantum critical behavior in a range of heavy fermion and strongly correlated electron materials,[45–47] yet the ability to spectroscopically trace the electronic structure across a field-tuned quantum phase transition has remained elusive. Given the sensitivity of the energy of the vHs in $Sr_2RuO_4$ to uniaxial strain,[6] we expect that the magnetic field at which the Lifshitz transition is extrapolated to occur here can be reduced substantially by combining uniaxial strain with magnetic field. This creates the opportunity to spectroscopically verify the role of quantum fluctuations across a magnetic field-tuned Lifshitz transition through a detailed study of the line shape of the vHs as it is tuned across the Fermi energy. The sensitivity of the electronic structure of ruthenates to tiny structural modifications found here shows opportunities for tailoring correlated electronic phases in 2D and exploring their physics for novel electronic devices.

## Supporting Information

Supporting Information is available from the Wiley Online Library or from the author.

## Acknowledgements

C.A.M. and L.C.R. contributed equally to this work. The authors thank J. Betouras, S. Grigera, T. Hanaguri, P. Hirschfeld, C. Hooley, P. King, A. Kreisel and S. Simon for fruitful discussions and David Miller for TEM characterization of the samples. C.A.M. acknowledges funding from EPSRC through EP/L015110/1, LCR from the Royal Commission for the Exhibition of 1851, A.W.R. from EPSRC through EP/P024564/1, P.W. from EPSRC through EP/R031924/1, and C.M.Y. and P.W. through EP/S005005/1. V.G., R.F., R.B., A.G., A.V. and P.W. acknowledge support from the Bilateral Project "Atomic-scale imaging of the superconducting condensate in the putative triplet superconductor $Sr_2RuO_4$: a platform for topological quantum computations?" in a joint Royal Society of Edinburgh and CNR Bilateral Scheme CUP B56C18003920005.

## Conflict of Interest

The authors declare no conflict of interest.

## Data Availability Statement

The data that support the findings of this study are openly available at the St Andrews University Research Portal at https://doi.org/10.17630/fc3776f6-dcd8-4624-a497-7ef8e770bb92.

## Keywords

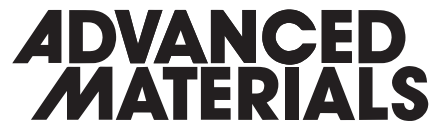

# Supporting Information

# Magnetic-Field Tunable Intertwined Checkerboard Charge Order and Nematicity in the Surface Layer of $Sr_2RuO_4$

*Carolina A. Marques, Luke C. Rhodes, Rosalba Fittipaldi, Veronica Granata, Chi Ming Yim, Renato Buzio, Andrea Gerbi, Antonio Vecchione, Andreas W. Rost, and Peter Wahl**

# Supplementary Material for

## Magnetic-field tunable intertwined checkerboard charge order and nematicity in the surface layer of $Sr_2RuO_4$

Carolina A. Marques,[1,*] Luke C. Rhodes,[1,*] Rosalba Fittipaldi,[2]
Veronica Granata,[3] Chi Ming Yim,[1] Renato Buzio,[4] Andrea Gerbi,[4]
Antonio Vecchione,[2] Andreas W. Rost,[1,5] and Peter Wahl[1,†]

[1]*School of Physics and Astronomy,*
*University of St Andrews, North Haugh,*
*St Andrews, KY16 9SS, United Kingdom*

[2]*CNR-SPIN, UOS Salerno, Via Giovanni Paolo II 132, Fisciano, I-84084, Italy.*

[3]*Dipartimento di Fisica "E. R. Caianiello",*
*Università di Salerno, I-84084 Fisciano, Salerno, Italy.*

[4]*CNR-SPIN, Corso F.M. Perrone 24, Genova, 16152, Italy*

[5]*Max-Planck-Institute for Solid State Research,*
*Heisenbergstr. 1, 70569 Stuttgart, Germany*

**This PDF file includes:**

Materials and Methods (S1)

Bias dependence of the checkerboard charge order in topographic STM images (S2)

Differential conductance maps (S3)

Tight-binding model (S4)

Analysis of magnetic-field dependent tunneling spectra (S5)

Figs. S1 to S9

*These authors contributed equally.

†correspondence to: wahl@st-andrews.ac.uk

## S1. MATERIALS AND METHODS

### A. Single crystal growth.

The $Sr_2RuO_4$ crystals used in this work were grown by the floating-zone technique with Ru self-flux, using a commercial image furnace with double-elliptical mirrors and two 2.0kW halogen lamps (*S1*). Morphological and elemental characterization of the crystals was carried out using a Zeiss Leo EVO 50 scanning electron microscope (SEM) equipped with an Oxford INCA Energy 300 energy dispersive X-ray spectroscopy (EDS) system. The structure and crystalline quality of the samples were assessed by a high-resolution X-ray diffractometer (Panalytical, X Pert MRD), with a Cu K-$\alpha$ source.

### B. Characterization

#### *1. Transmission electron microscopy (TEM)*

To verify the crystal quality in the surface region following cleavage of the sample in the STM, we have performed Transmission Electron Microscopy on a sample on which STM measurements had been carried out. The sample for STEM analysis was prepared by conventional gallium focused ion beam (FIB) milling using an FEI Scios focused ion beam scanning electron microscope (FIBSEM) equipped with an EDAX Hikari Super electron back-scattered diffraction (EBSD) detector. The orientation of the sample was determined by EBSD prior to milling, to cut a lamella in the [010] plane. High angle annular dark field (HAADF) images were recorded using a probe corrected FEI Themis 200 scanning/transmission electron microscope operated at 200kV.

Figure S1(a) shows a TEM image of a cross section of the sample after cleaving. It shows a uniform phase along the $c$-axis on the scale of $\approx$ 80nm extending up to the surface. Figure S1(b) shows a zoom in with atomic resolution. It shows the expected stacking for $Sr_2RuO_4$, as evidenced by the inset, where both strontium and ruthenium atoms are visible, with the Ru atoms appearing with a slightly higher intensity. Oxygen atoms are not visible. The TEM image shows no evidence of inclusions of other members of the Ruddlesden-Popper series (i.e. $Sr_{n+1}Ru_nO_{3n+1}$ with $n > 1$).

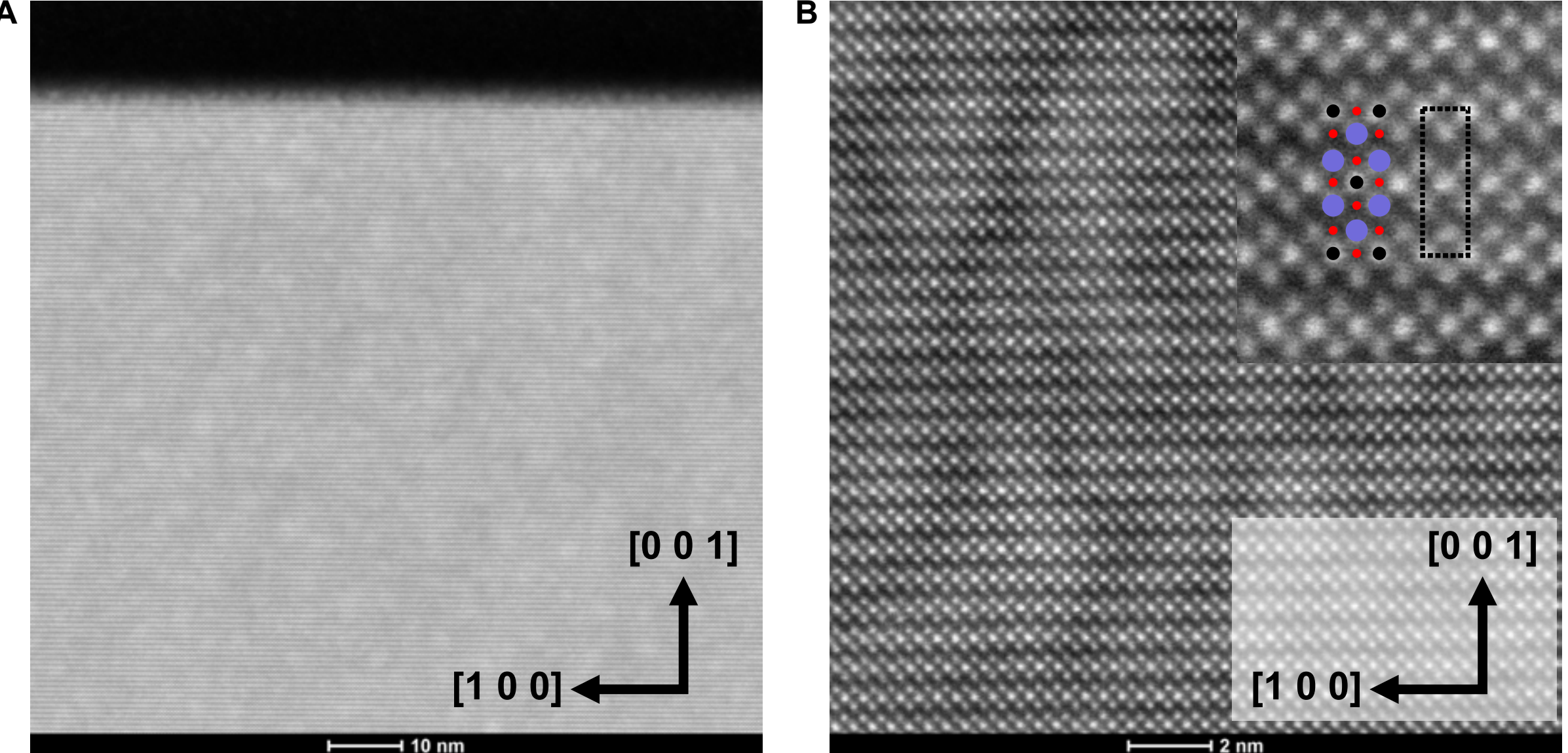

FIG. S1: TEM images along the b-axis. (a) Image of the cross-section of a sample after STM measurements have been performed on its surface, demonstrating uniformity up to the surface. (b) High-resolution image showing atomic resolution. The inset shows a zoom in of the image, with the stacking expected for $Sr_2RuO_4$. Superimposed to the TEM image is the atomic structure, with spheres indicating strontium (purple), ruthenium (black) and oxygen (red) atoms. The black dashed line indicates one unit cell of $Sr_2RuO_4$.

### 2. *Resistance measurements*

Transport measurements were performed using a four probe technique with a $^3$He refrigerator. The resistance as a function of temperature (Fig. S2) shows a superconducting transition at $T_c = 1.5$K, in agreement with data reported in literature for good quality crystals(*S2*).

We determine a residual-resistance ratio (RRR) $\lim_{T\to 0\mathrm{K}} \frac{R(300\mathrm{K})}{R(T)} \approx 666$, extrapolated to 0K from fitting the resistance above $T_c$ to $R(T) = R_{\mathrm{res}} + AT^2$ (Figure S2(a)). This value is comparable to that of high-purity crystals reported in the literature(*S3*).

### 3. *Scanning tunneling microscopy (STM)*

Experiments were performed in a home-built ultra-low temperature STM operating in a dilution refrigerator.(*S4*) Samples were prepared by *in-situ* cleaving at low temperatures ($\sim$ 20K) in cryogenic vacuum. We used STM tips cut from PtIr wire, and prepared them *in-situ* by field emission on a Au(111) single crystal. Bias voltages were applied to the sample, with the tip at virtual ground.

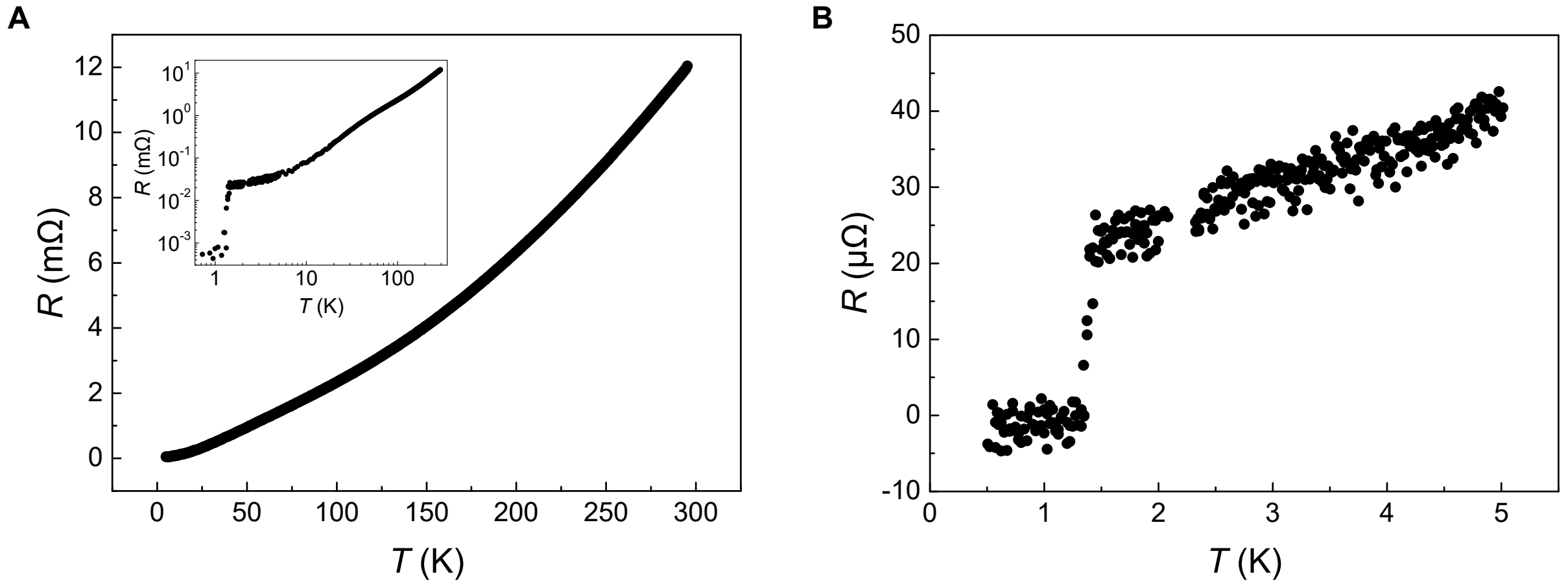

FIG. S2: **Resistance as a function of temperature of $Sr_2RuO_4$ with current in the a-b plane.** (a) Resistance up until room temperature. The inset shows the resistance as function of temperature in log-log scale (b) Low temperature resistance down to 0.5 K showing the superconducting transition at 1.5 K.

Spectroscopic measurements were performed using a lock-in amplifier to measure differential conductance $g(V)$. The bias voltage $V$ was modulated at a frequency of $\nu = 397$Hz, the lock-in modulation $V_{\mathrm{L}}$, specified as amplitude, is given in figure captions where applicable. The bias and current setpoints, $V_{\mathrm{set}}$ and $I_{\mathrm{set}}$, are indicated for both topographies and spectroscopic maps.

After cleaving and inserting the samples into the STM head, we have always observed the reconstructed SrO termination as observed in previous reports(*S5–S8*) and as can be inferred from the appearance of the defects and the additional Fourier peaks due to the checkerboard charge order.

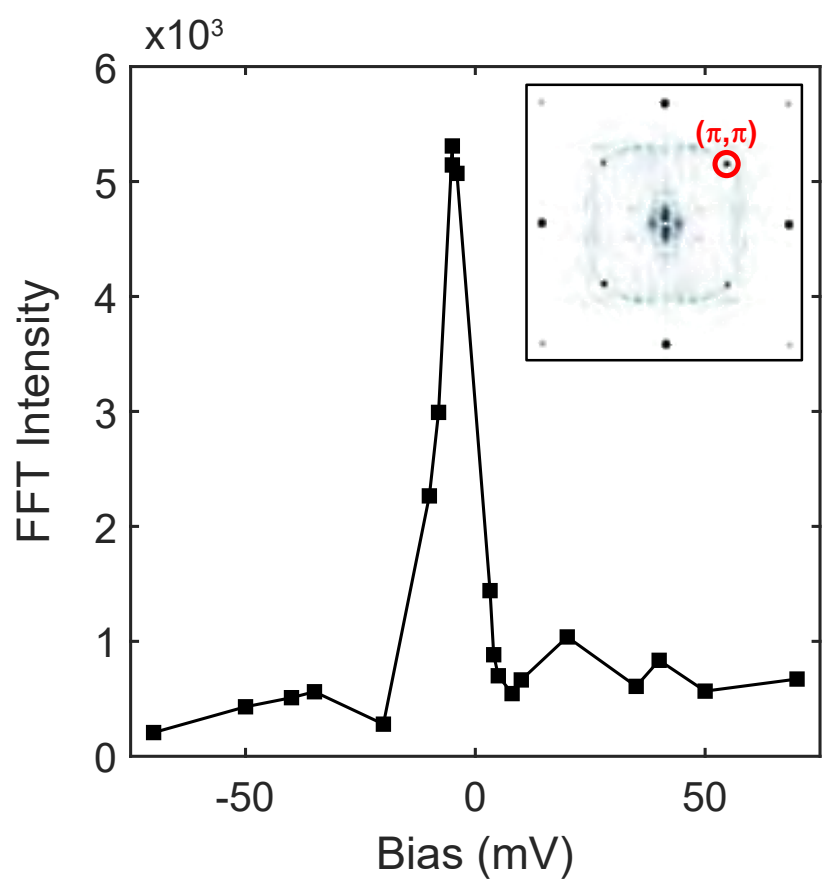

FIG. S3: **Bias dependence of the checkerboard charge order in topographic STM images.** Intensity of the peak in the Fourier transformation $\tilde{z}(\mathbf{q})$ of topographic images $z(\mathbf{r})$ corresponding to the checkerboard charge order at $\mathbf{q}_{\mathrm{ckb}} = (\pi, \pi)$ for bias voltages between $+/-70$mV. The inset shows a typical Fourier transformation taken with $-5$mV, where the peak at $\mathbf{q}_{\mathrm{ckb}}$ is indicated by a red circle. This data was taken at 2K and the tunneling resistance was kept at 100MΩ for all measurements.

## S2. BIAS DEPENDENCE OF THE CHECKERBOARD CHARGE ORDER IN TOPOGRAPHIC STM IMAGES

The appearance of the checkerboard modulation in topographies is bias dependent, as seen in Figures 2(a) and (b) of the main manuscript. Here we show the bias dependence of the appearance of the checkerboard charge order in the voltage range between $-70$mV and $+70$mV. We plot the intensity $\tilde{z}(\mathbf{q}_{\mathrm{ckb}})$ of the peak in the Fourier transformation associated with the checkerboard order ($\mathbf{q}_{\mathrm{ckb}} = (\pi, \pi)$) as a function of applied bias voltage $V$ in Figure S3. The intensity of the checkerboard charge order has a sharp maximum around $-5$mV, with its intensity decreasing by a factor of 25 at $-70$mV and of 8 at $+70$mV. Using this information, for spectroscopic maps shown here, we have chosen a tunneling setpoint at small positive bias voltages where the checkerboard charge order is weak to minimize the setpoint effect.

## S3. DIFFERENTIAL CONDUCTANCE MAPS

In this section, we show the real space conductance maps underpinning the data shown in Fig. 3(d) and how it has been processed.

### A. Real space maps

The topographic image acquired simultaneously with the map taken at 2K is shown in Figure S4(a) and the corresponding real space differential conductance map, $g(\mathbf{r}, V) = \mathrm{d}I/\mathrm{d}V(\mathbf{r}, V)$, for $V = -3.4\mathrm{mV}$ in Figure S4(b). The checkerboard charge order is clearly visible. In addition quasiparticle interference (QPI) effects are observed around the defect seen in the topography.

### B. Processing of differential conductance maps

In Figure S4 we show images following each step of the data processing. The raw Fourier transformation of the $g(\mathbf{r}, V)$ map is displayed in Fig. S4(c). It shows clear signatures of the atomic peaks at $(\pm 2\pi, 0)$ and $(0, \pm 2\pi)$ (in units of $1/a$, where $a$ is the lattice constant of the bulk tetragonal unit cell), as well as of the checkerboard charge order at $(\pm\pi, \pm\pi)$ and $(\pm\pi, \mp\pi)$. There are in addition weak higher order peaks. The Fourier transformation is first corrected for any linear drift by mapping the atomic peaks by a linear transformation onto a perfect square (Fig. S4(d)). Next, the resulting images are mirror-symmetrized along the horizontal or vertical direction (Fig. S4(e)). Finally, to suppress the high intensity at the centre of the image due to the distribution of defects, the image is multiplied by 1 minus a two-dimensional Gaussian function with a standard deviation of 4 pixels (Fig. S4(f)).

This procedure was implemented for all layers of the map. Fig. S5 shows the Fourier transformation after processing of the same $g(\mathbf{r}, V)$ map at different energies, between $-8.2\mathrm{mV}$ and $+7.8\mathrm{mV}$. The intensity of the $(\pm\pi, \pm\pi)$ peaks is strongly energy dependent, showing maximum intensity at $-3.4\mathrm{mV}$ (highlighted by a red box). Additionally, we observe an anisotropy in the intensities of the $(2\pi,0)$ and $(0,2\pi)$ peaks, as is indicated in Fig. S5 by the circles drawn with solid red and dashed white lines. It can be seen that the high intensity switches from one direction to the other with energy. This reflects the atomic scale symmetry

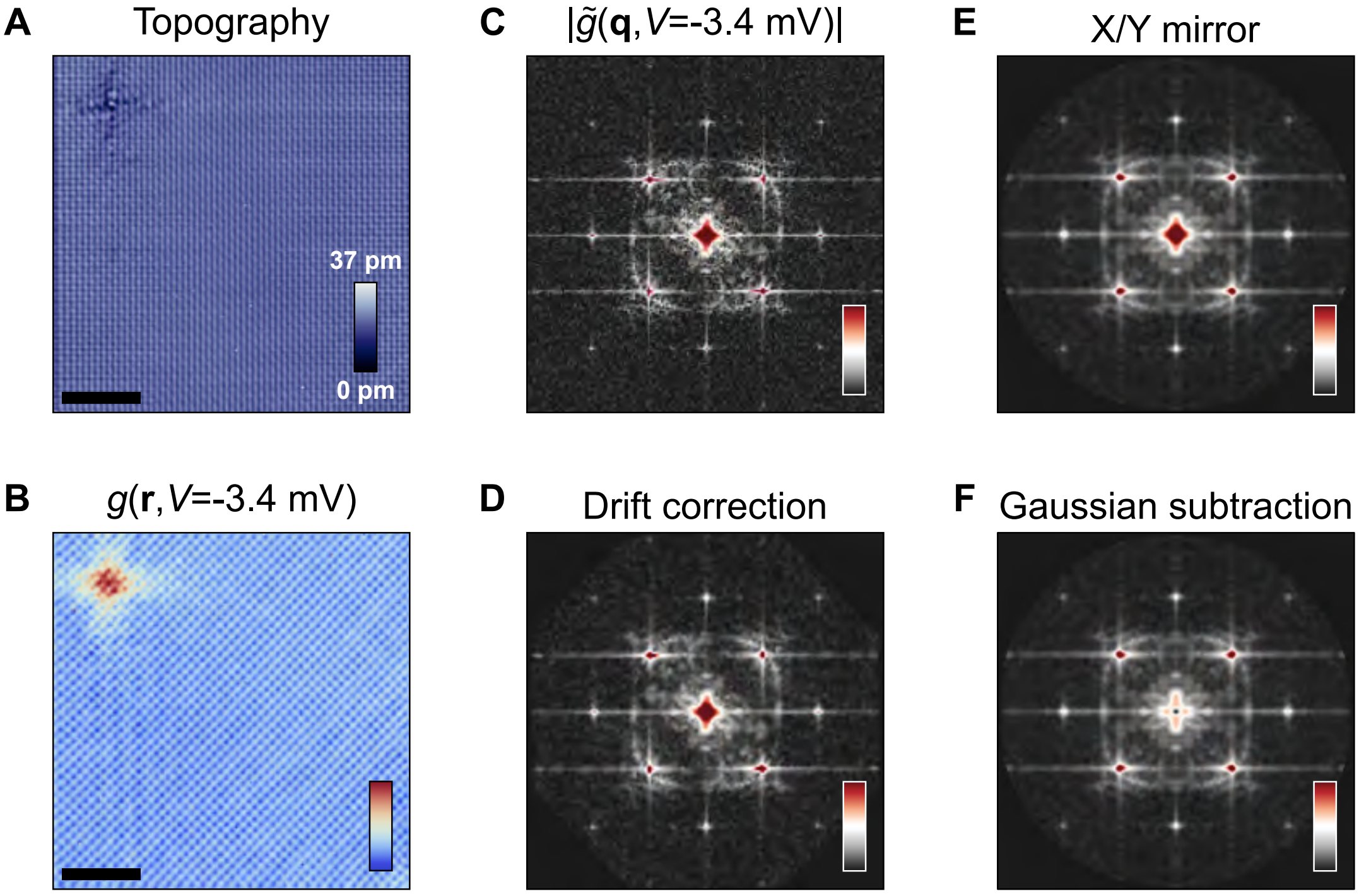

FIG. S4: **Processing of differential conductance maps.** (a) Topography acquired simultaneously with map, showing one defect in the top left corner. Scale bar: 5nm ($V_{\mathrm{set}} = 7.8$mV, $I_{\mathrm{set}} = 225$pA). (b) Real space $g(\mathbf{r}, V)$ at $V = -3.4$ mV. (c) Absolute value of its Fourier transformation, $|\tilde{g}(\mathbf{q}, V)|$, (d) after drift correction, (e) after symmetrizing along the horizontal (or vertical) direction, and (f) after subtraction of a gaussian function with a width of 4 pixels at the centre of the image ($V_{\mathrm{L}} = 800\mu$V, $T = 2$K, $B_z = 0$T).

breaking shown in Fig. 4(b) and (c) in the main text and is linked to the nematicity.

### C. Phase-referenced Fourier transformation

In figs. 3(d) and 6(c) of the main text, we show phase-referenced Fourier transformations to deduce the characteristic energy scale of the checkerboard charge order and the relative phase at positive and negative bias voltages. The Fourier transformation $\tilde{g}(\mathbf{q}, V)$ of a differential conductance map $\tilde{g}(\mathbf{r}, V)$ which shows the atomic lattice exhibits two pairs of Bragg peaks at $\mathbf{q}_{\mathrm{at}} = (\pm 2\pi, 0)$ and $(0, \pm 2\pi)$, indicated by a black circle in Figure S6(a). The checkerboard order shown in Figure 3(b), (c) of the main text has a wavelength that is $\sqrt{2}$ times larger, with the unit vector 45° rotated. The periodicity due to the checkerboard charge order shows up as four peaks at $\mathbf{q}_{CKB} = (\pm\pi, \pm\pi)$ and $\mathbf{q}_{CKB} = (\pm\pi, \mp\pi)$, one of which is indicated by the red circle in Fig. S6(a). The amplitude of this checkerboard order is reflected in the intensity of the Fourier peak, $|\tilde{g}(\mathbf{q}_{CKB}, V)|$, as shown in Fig. S5. We can determine the characteristic energy scale of the checkerboard modulation as a function of energy by plotting, $|\tilde{g}(\mathbf{q}_{CKB}, V)|$ as a function of applied bias $V$, however losing the phase information $\phi(\mathbf{q}, V)$ contained in the Fourier transformation,

$$\tilde{g}(\mathbf{q}, V) = \sqrt{\frac{\Delta x \Delta y}{N_x N_y}} \mid \tilde{g}(\mathbf{q}, V) \mid e^{i\phi(\mathbf{q}, V)}, \tag{S1}$$

where $\Delta x$ and $\Delta y$ are the size of the map in along the $x$ and $y$ directions and $N_x$ and $N_y$ are the number of pixels in each direction. While analyzing the phase $\phi(\mathbf{q}, V)$ itself is possible, it suffers from an arbitrary global phase factor. To remove this global phase factor, we use a phase-referenced Fourier transformation (PR-FT)(*S9*)

$$\tilde{g}^{\mathrm{R}}(\mathbf{q}, V) = \frac{\tilde{g}(\mathbf{q}, V)}{e^{i\phi(\mathbf{q}, V_0)}} = \sqrt{\frac{\Delta x \Delta y}{N_x N_y}} \mid \tilde{g}(\mathbf{q}, V) \mid e^{i(\phi(\mathbf{q}, V) - \phi(\mathbf{q}, V_0))}. \tag{S2}$$

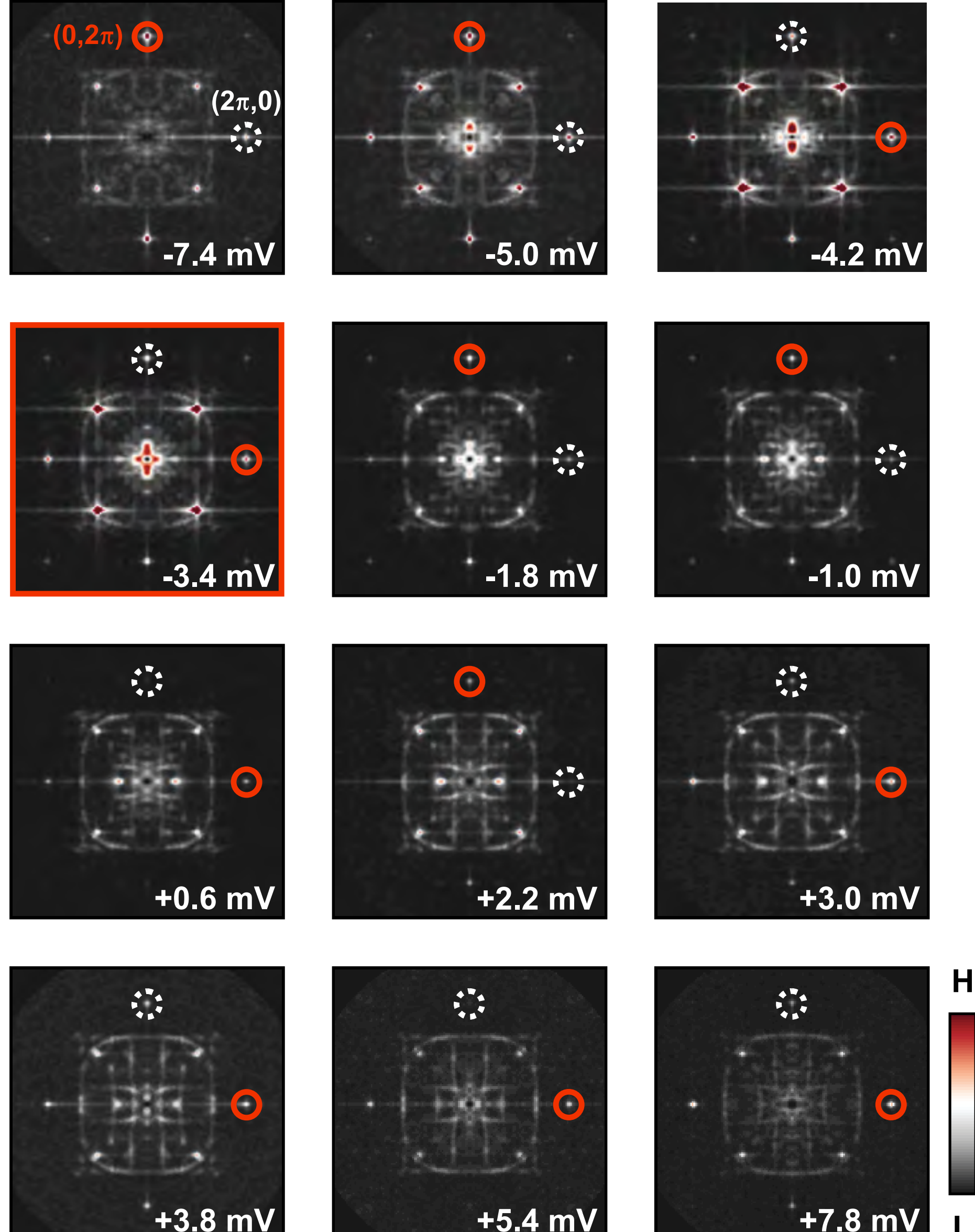

FIG. S5: **Energy layers from the $\tilde{g}(\mathbf{q}, V)$ map** shown in Fig. 3(d) of the main text and Fig. S4 for the data at 2K. The intensity of the peak at $\mathbf{q}_{\text{CKB}} = (\pi, \pi)$ due to the checkerboard charge order is strongly energy dependent. The layer corresponding to the energy of the vHs is highlighted by a red box, where the peaks at $\mathbf{q}_{\text{CKB}}$ exhibit the highest intensity. Additionally, the intensity of the Bragg peaks $\mathbf{q}_{\mathbf{at}} = (0, \pm 2\pi)$ and $(\pm 2\pi, 0)$ switches with energy due to the nematicity (cf. fig. 4(b), (c) of the main text). For each layer, the higher intensity peak is highlighted by a red circle, and the lower intensity peak with a dashed white circle. The range of the color bar is the same for all images.

In this PR-FT, the phase at each $\mathbf{q}$-vector is referenced to the phase at a specific energy $V_0$, removing the global phase factor. This allows tracking of the change in phase as a function of energy relative to that layer by simply plotting the real part of the PR-FT,

$$Re\left[\tilde{g}^{\mathrm{R}}(\mathbf{q}, V)\right] = \sqrt{\frac{\Delta x \Delta y}{N_x N_y}} \mid \tilde{g}(\mathbf{q}, V) \mid \cos\left(\phi(\mathbf{q}, V) - \phi(\mathbf{q}, V_0)\right). \tag{S3}$$

Here, we reference the phase to the map layer at $V_0 = -3.4$mV. The real part of the PR-FT images, Figure S6(c), $\mathrm{Re}[\tilde{g}^{\mathrm{R}}((\pi, \pi), V)]$ allows us to determine if there is a phase shift between the checkerboard order at different energies as well as the relative amplitude. A phase reversal means a change in sign. At the reference energy, $V_0$, $\mathrm{Re}[\tilde{g}^{\mathrm{R}}((\pi, \pi), V_0]$ will be positive by definition. Fig. S6(c) shows $\mathrm{Re}[\tilde{g}^{\mathrm{R}}((\pi, \pi), V = +3\mathrm{mV}]$, where the peaks at $\mathbf{q}_{\mathrm{CKB}}$ appear with a negative sign, evidencing a phase shift with respect to the charge modulation at $V_0 = -3.4$mV. Figure 3(d) of the main text shows that the checkerboard order appears predominantly at $V = -3.5$mV and $V = 3.5$mV with opposite phase.

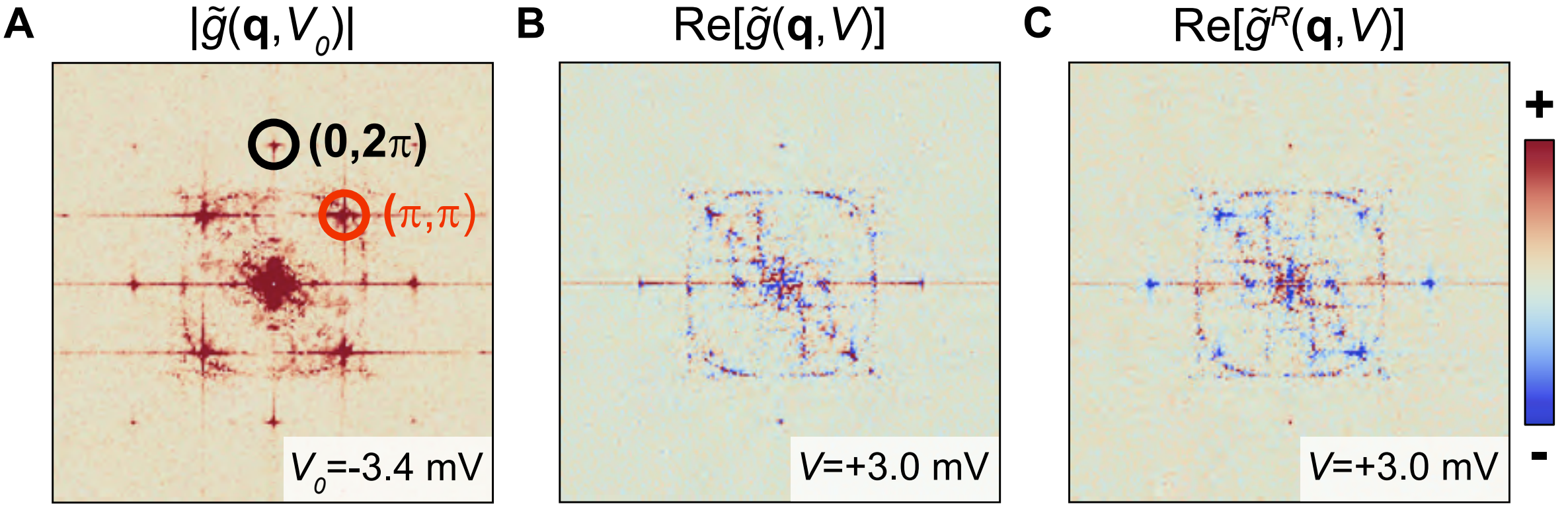

FIG. S6: **Phase-referenced Fourier transformation.** (a) Amplitude $|\tilde{g}(\mathbf{q}, V_0 = -3.4\mathrm{mV})|$ of the Fourier transformation. (b) Real part $\mathrm{Re}[\tilde{g}(\mathbf{q}, V = +3.0\mathrm{mV})]$, it can be seen that the peaks at $(\pi, \pi)$ and $(-\pi, \pi)$ have opposite sign. (c) Real part of the phase-referenced Fourier transformation (PR-FT), $\mathrm{Re}[\tilde{g}^{\mathrm{R}}(\mathbf{q}, V = +3.0\mathrm{mV})]$, showing the reversed sign of the peaks at $(\pi, \pi)$ and $(-\pi, \pi)$ compared to panel A (map parameters as in fig. S4).

## S4. TIGHT-BINDING MODEL

The tight-binding model for the surface of $Sr_2RuO_4$ may be defined, in an analogous manner to that of the bulk, via hopping associated with the three bands derived from the ruthenium $t_{2g}$ orbitals (*S10*),

$$H^\sigma(\mathbf{k}) = \begin{pmatrix} E_{xz}(\mathbf{k}) & \gamma(\mathbf{k}) - \sigma i\eta & i\eta \\ \gamma(\mathbf{k}) + \sigma i\eta & E_{yz}(\mathbf{k}) & -\sigma\eta \\ -i\eta & -\sigma\eta & E_{xy}(\mathbf{k}) \end{pmatrix}. \tag{S4}$$

This Hamiltonian includes nearest neighbour hoppings between the $d_{xz}$ and $d_{yz}$ orbitals as well as up to third nearest neighbour hoppings between $d_{xy}$ orbitals,

$$
\begin{aligned}
E_{xz}(\mathbf{k}) &= -2t_1\cos(k_x) - 2t_2\cos(k_y) - \mu, \\
E_{yz}(\mathbf{k}) &= -2t_2\cos(k_x) - 2t_1\cos(k_y) - \mu, \\
E_{xy}(\mathbf{k}) &= -2t_3(\cos(k_x) + \cos(k_y)) - 4t_4\cos(k_x)\cos(k_y) - 2t_5(\cos(2k_x) + \cos(2k_y)) - \mu_c.
\end{aligned}
\tag{S5}
$$

The off diagonal term, $\gamma(\mathbf{k})$, describes inter-orbital hopping between the degenerate $d_{xz}$ and $d_{yz}$ states and is written as $\gamma(\mathbf{k}) = -4t_{\text{inter}}\sin(k_x)\sin(k_y)$. The rest of the off-diagonal terms arise from the spin-orbit interaction, where $\eta$ is the spin-orbit coupling constant and $\sigma$ is defined as $+1$ for up spins ($\uparrow$) and $-1$ for down spins ($\downarrow$).

Using this basis, a tight binding model for the electronic structure can be defined via the Hamiltonian

$$
H_{\text{Ru}}(\mathbf{k}) = \begin{pmatrix} H^{\uparrow}(\mathbf{k}) & 0 \\ 0 & H^{\downarrow}(\mathbf{k}) \end{pmatrix},
\tag{S6}
$$

representing now a $6 \times 6$ matrix. For the surface, we must account for the doubling of the unit cell due to the additional octahedral rotation. This surface tight binding model is then given by the $12 \times 12$ matrix

$$
H_{\text{surf}}(\mathbf{k}) = \begin{pmatrix} H_{\text{Ru}}(\mathbf{k}) & 0 \\ 0 & H_{\text{Ru}}(\mathbf{k}+\mathbf{Q}) \end{pmatrix},
\tag{S7}
$$

with $\mathbf{Q} = (\pi, \pi)$.

For the bulk Fermi surface, presented in Fig. 1(b) of the main text, we calculate the Eigenvalues using Eq. S6 with the hopping parameters from Ref. *S10.*

$$
\begin{array}{lllll}
t_1 = 0.15\text{eV} & t_2 = 0.1t_1 & t_3 = 0.8t_1 & t_4 = 0.3t_1 & t_5 = 0\text{eV} \\
t_{\text{inter}} = 0.01t_1 & \mu = 1.0t_1 & \mu_c = 1.1t_1 & \eta = 0.1t_1. &
\end{array}
$$

Here $t_1$ is set to 150 meV and all other parameters are defined relative to $t_1$. For the Fermi surface presented in Fig. 1(d) we change $t_5 = 0.1t_1$, $\mu_c = 0.75t_1$ and $\mu = 0.82t_1$ in order to describe a system with a vHs located just below the Fermi level.

For the tight-binding description of the surface electronic structure, presented in Fig. 5, we use the hopping parameters

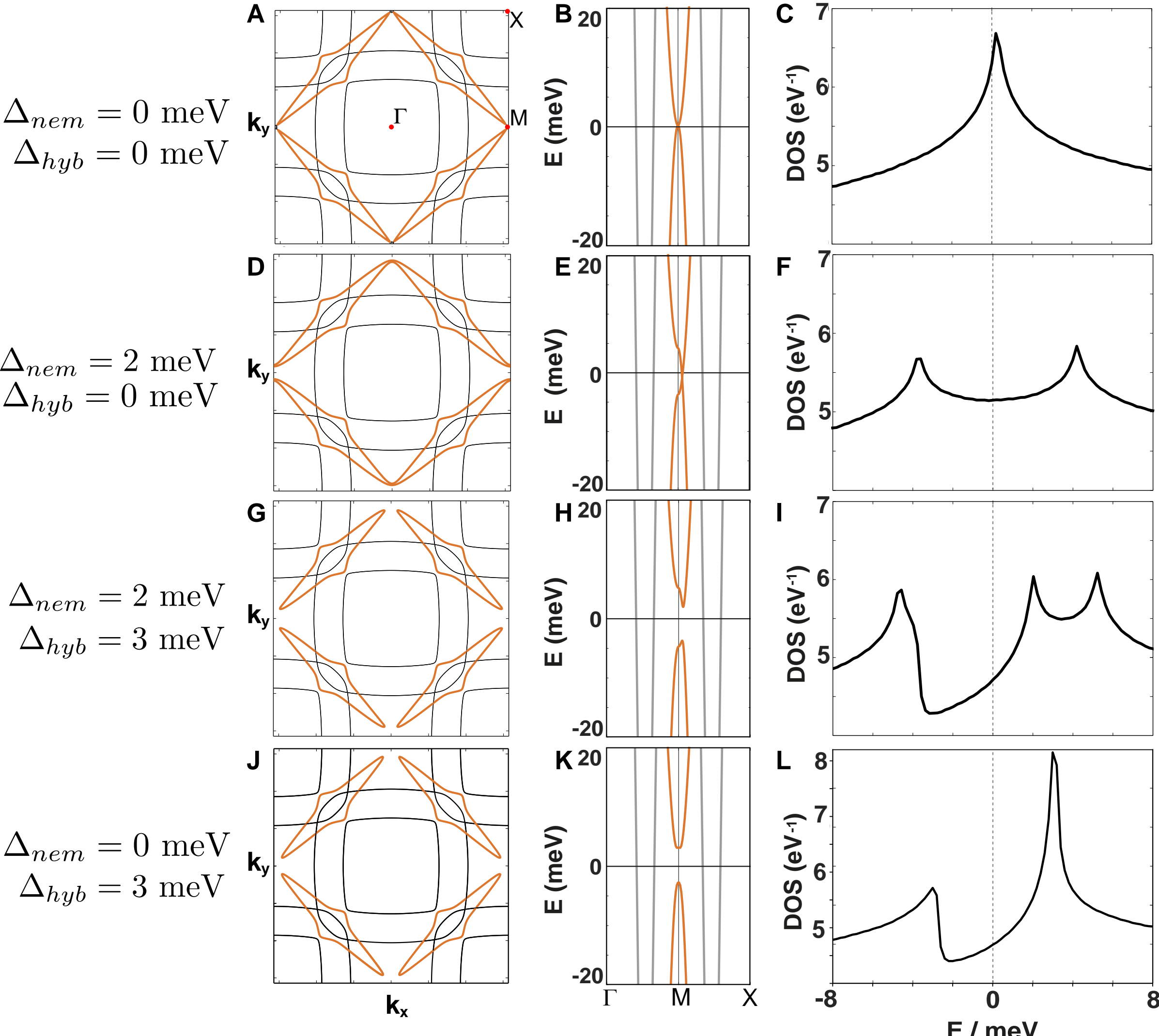

FIG. S7: Tight-binding model of the surface of $Sr_2RuO_4$. (a)-(c) Fermi surface, band dispersion along Γ-M-X and density of states between ±8meV for a model without nematicity ($\Delta_{\mathrm{nem}} = 0$meV and hybridization of bands due to the reconstruction $\Delta_{\mathrm{hyb}} = 0$meV. (d)-(f) Equivalent plots with nematicity included, $\Delta_{\mathrm{nem}} = 2$meV, but no hybridization between the bands due to the reconstruction $\Delta_{\mathrm{hyb}} = 0$meV. (g)-(i) Equivalent plots with nematicity $\Delta_{\mathrm{nem}} = 2$meV (as before) and a non-zero hybridization $\Delta_{\mathrm{hyb}} = 3$meV. (j)-(l) Equivalent plots with finite $\Delta_{\mathrm{hyb}}$ only.

$$t_1 = 0.15\mathrm{eV} \qquad t_2 = 0.1t_1 \qquad t_3 = 0.8t_1 \qquad t_4 = 0.3t_1 \qquad t_5 = 0.095t_1$$

$$t_{\mathrm{inter}} = 0.01t_1 \qquad \mu = 0.75t_1 \qquad \mu_c = 0.812t_1 \qquad \eta = 0.1t_1.$$

This places the vHs just above the Fermi level and slightly decreases the Fermi wave vector, $k_F$, of the $d_{xz}$ and $d_{yz}$ bands as suggested by ARPES measurements (*S11*). We note that the shape of the $d_{xy}$ related pockets observed in both ARPES and DFT are slightly larger than in our model, however this difference does not affect the conclusions drawn here. We then introduce a hybridisation between the two Ru sites, $\Delta_{\text{hyb}} = 3\text{meV}$, as off-diagonal elements in Eq. S7,

$$H_{\text{surf}}(\mathbf{k}) = \begin{pmatrix} H_{\text{Ru}}(\mathbf{k}) & \Delta_{\text{hyb}}\hat{I} \\ \Delta^*_{\text{hyb}}\hat{I} & H_{\text{Ru}}(\mathbf{k}+\mathbf{Q}) \end{pmatrix}, \tag{S8}$$

and include a phenomenological $C_4$ symmetry breaking term $\Delta_{\text{nem}}(\mathbf{k}) = \delta_{\text{nem}}(\cos(k_x) - \cos(k_y))$ specifically to the $d_{xy}$ orbital in $H_{\text{Ru}}$, with $\delta_{\text{nem}} = 2\text{meV}$, to produce the full Hamiltonian defined in Eq. 1 of the main text. A similar nematic term has been discussed for $Sr_3Ru_2O_7$ previously.(*S12*)

$$H_{\text{surf}}(\mathbf{k}) = \begin{pmatrix} H_{\text{Ru}}(\mathbf{k}) + \Delta_{\text{nem}}(\mathbf{k})\hat{I}_{xy} & \Delta_{\text{hyb}}\hat{I} \\ \Delta^*_{\text{hyb}}\hat{I} & H_{\text{Ru}}(\mathbf{k}+\mathbf{Q}) + \Delta_{\text{nem}}(\mathbf{k}+\mathbf{Q})\hat{I}_{xy} \end{pmatrix}. \tag{S9}$$

The density of states presented in Fig. 5(c) has been calculated via

$$N_0(\omega) = -\frac{1}{\pi}\text{Tr}\big[\text{Im}\big[\sum_{\mathbf{k}} G(\mathbf{k},\omega)\big]\big]. \tag{S10}$$

Here, $G(\mathbf{k},\omega)$ is the Green's function defined as $G(\mathbf{k},\omega) = \frac{1}{\omega - H(\mathbf{k}) + i\Gamma}$, where $\omega$ is the energy and $\Gamma$ a broadening parameter. We use a $\mathbf{k}$-grid of 4096x4096 lattice points and $\Gamma = 0.1\text{meV}$.

Fig. S7 shows the Fermi surface, band structure and DOS given by the tight-binding model described above. Fig. S7(a)-(c) show the case where only the doubling of the unit cell is taken into account (Eq. S7). The vHs was put above, but very close to, $E_{\text{F}}$ and it appears as a sharp peak with logarithmic divergence, cut off by the broadening parameter $\Gamma$. Fig. S7(d)-(f) show the case were only the $C_4$-symmetry breaking term is included. The $d_{xy}$ band becomes $C_2$-symmetric, and the vHs splits into two peaks, one above $E_{\text{F}}$ and another below, although no gap opens around the Fermi energy. Fig. S7(g)-(i) show the case of eq. S9, where both the nematic term and the hybridization potential are included. A gap is opened between the $d_{xy}$ bands, creating four vHs. The calculated DOS reproduces the measured differential conductance spectrum, as shown in Fig. 5 of the main text.

In Fig. S8, we present the DOS calculated using the surface tight-binding model in the presence of a magnetic field. To simulate this, we introduce a Zeeman splitting term to the

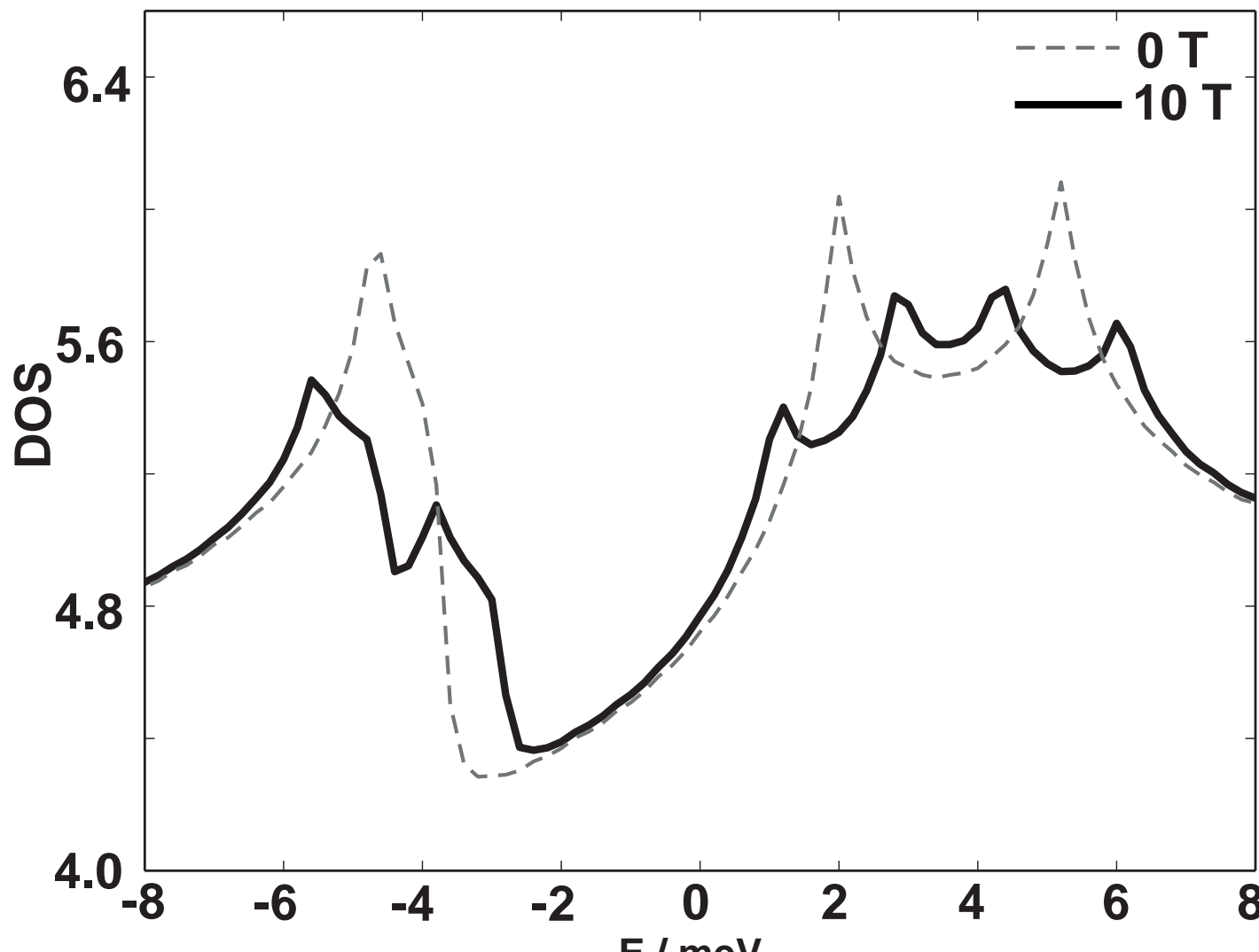

FIG. S8: DOS calculated using the surface tight binding model with a Zeeman splitting term due to a magnetic field $B_z = 10$T. The DOS calculated with $B_z = 0$T is also shown for comparison.

Hamiltonian

$$H_{\mathrm{Field}}(\mathbf{k}) = H_{\mathrm{surf}}(\mathbf{k}) + \frac{\sigma}{2} g^* \mu_{\mathrm{B}} B_z \hat{I}. \tag{S11}$$

Here $\sigma = +1$ for up spin states and $-1$ for down spin states. $g^*$ has been set to 3, as determined experimentally and discussed in the main text, $\mu_{\mathrm{B}} = 5.788 \cdot 10^{-5}$ eVT$^{-1}$ and $B_z$ is the magnetic field strength in the $z$-direction. The introduction of a magnetic field reduces the peak height of the vHs's.

## S5. ANALYSIS OF MAGNETIC-FIELD DEPENDENT TUNNELING SPECTRA

Each $g(\mathbf{r}, V)$ spectrum shown in Figure 6(a) of the main text is obtained from an average of 10 spectra. All spectra were acquired with the same setpoint conditions before turning off the feedback loop. To determine the energy of the van Hove singularity from the peak positions in the spectra, we first subtract a background from the spectra, and then fit the positions of the dominant peaks. To describe the background, we fit an arc tangent and a constant ($f(V) = a \cdot \arctan[(V - V_0)/\Gamma] + c$) to the background in the data at 13.4T to describe the gap edge at negative energies and subtract the resulting function $f(V)$ as a base line. For the data at other fields, we fit the values of $a$ and $c$ to the background using the same arctangent function, but keep $V_0$ and $\Gamma$ fixed at the values of the fit at 13.4T, Figure S9(a). To determine the energy of the van Hove singularity, we fit a Lorentzian to the peak in the spectrum, Figure S9(b).

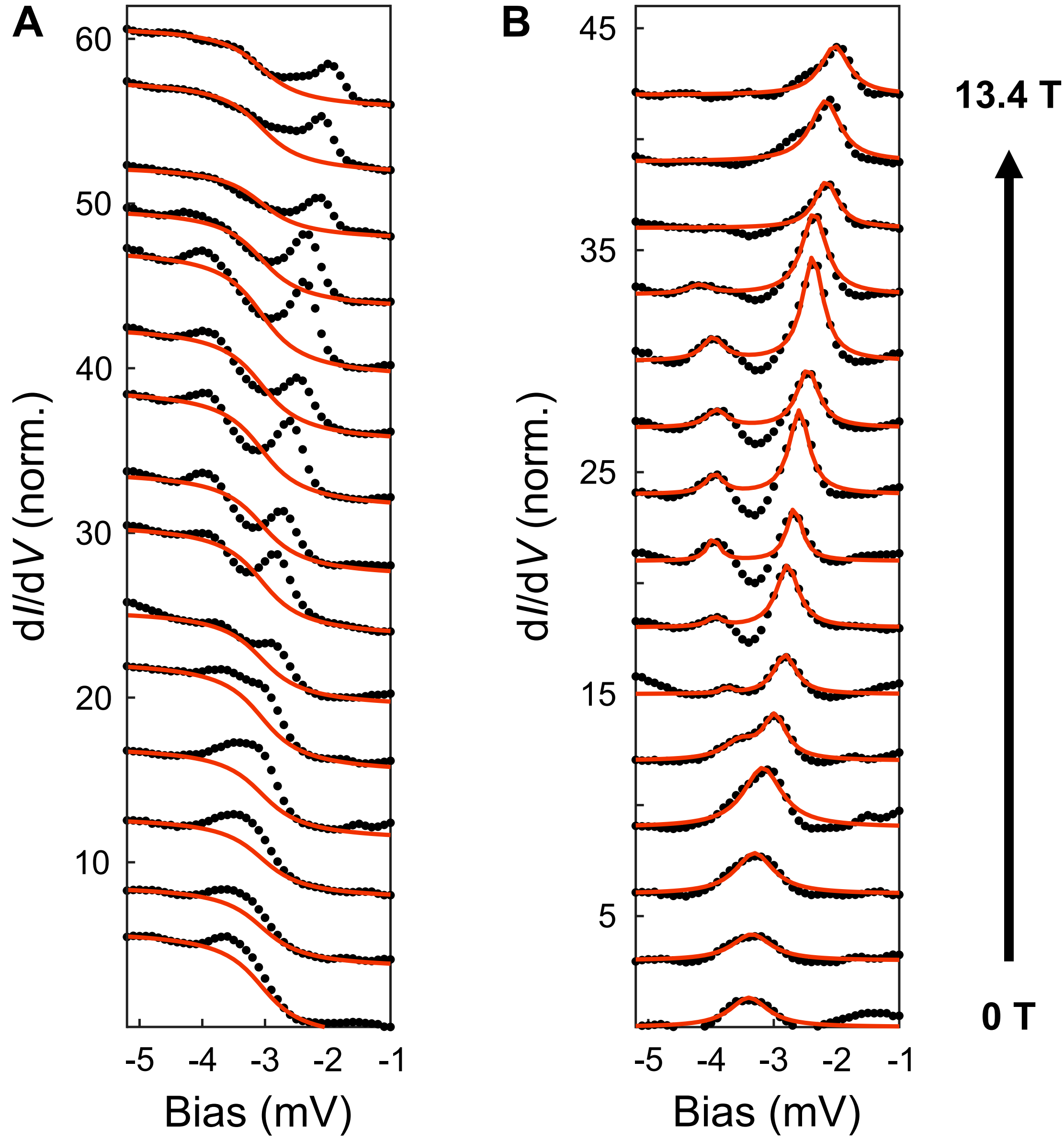

FIG. S9: **Determination of splitting of the vHs in magnetic field.** (a) Differential conductance spectra $g(\mathbf{r}, V)$ in magnetic fields of $B = 0 \ldots 13.4$T, with the background fit shown as red lines. (b) $g(\mathbf{r}, V)$ spectra after subtraction of the background. The red lines show the lorentzian fits to extract the peaks positions.